\documentclass[aps,prl,twocolumn,superscriptaddress,10pt]{revtex4-2}
\usepackage[utf8]{inputenc}
\usepackage[T1]{fontenc}
\usepackage{silence}
\usepackage{amsmath, amssymb}
\usepackage{graphicx}
\usepackage{mathrsfs}
\usepackage{bm}
\usepackage{xcolor}
\usepackage{csquotes}
\usepackage[colorlinks=true,allcolors=purple]{hyperref}

\definecolor{tab_blue}{HTML}{1F77B4}
\definecolor{tab_orange}{HTML}{FF7F0E}
\definecolor{tab_green}{HTML}{2CA02C}
\definecolor{tab_red}{HTML}{D62728}
\definecolor{tab_purple}{HTML}{9467BD}
\definecolor{tab_brown}{HTML}{8C564B}
\definecolor{tab_pink}{HTML}{E377C2}
\definecolor{tab_gray}{HTML}{7F7F7F}
\definecolor{tab_olive}{HTML}{BCBD22}
\definecolor{tab_cyan}{HTML}{17BECF}

\makeatletter
\def\figurename{Fig.}
\makeatletter
\def\fnum@figure{\textbf{\figurename~\thefigure}}
\makeatother

\def\TT{\mathsf{T}}
\def\dd{\text{d}}
\def\ii{\text{i}}
\def\ee{\text{e}}

\begin{document}

\title{Odd stochastic density functional theory}

\author{Yael Avni}
\affiliation{University of Chicago, James Franck Institute, 929 E 57th Street, Chicago, IL 60637}
\affiliation{Department of Physics of Complex Systems, Weizmann Institute of Science, 76100 Rehovot, Israel}

\author{Michel Fruchart}
\affiliation{Gulliver, ESPCI Paris, Université PSL, CNRS, 75005 Paris, France}

\author{David Martin}
\affiliation{LPTMC, CNRS UMR 7600, Université Pierre et Marie Curie, 75252 Paris, France}

\author{Tali Khain}
\affiliation{John A. Paulson School of Engineering and Applied Sciences, Harvard University, Cambridge, MA 02138, USA}

\author{Vincenzo Vitelli}
\affiliation{University of Chicago, James Franck Institute, 929 E 57th Street, Chicago, IL 60637}
\affiliation{Leinweber Institute for Theoretical Physics, University of Chicago, Chicago, IL 60637, USA}

\date{\today}

\begin{abstract}
Stochastic density functional theory (SDFT) provides a powerful framework for studying the dynamics of many-body systems both in and out of equilibrium. In its standard form, it can be viewed as a density-based formulation of coupled Langevin equations with white noise. 
Here, we extend SDFT to a class of systems, ranging from magnetized electrolytes to chiral active matter, that possess a \enquote{chiral overdamped limit} in which a relaxation timescale and an oscillation period are simultaneously taken to zero.
We derive this formulation, which we dub “odd SDFT”, by taking the overdamped limit of a fluctuating hydrodynamic description of underdamped particles subject to a Lorentz force. We establish the validity and utility of the resulting framework by calculating the collective mobility tensor of interacting particles using both direct linear response and a Green-Kubo relation. We then apply the theory to the Hall conductivity of electrolytes in magnetic fields. 
Going beyond mean-field theory and using two distinct Green-Kubo relations, we derive the full conductivity tensor, extending the relaxation part of the Debye-H{\"u}ckel-Onsager theory to the case of finite magnetic fields.
\end{abstract}

\maketitle

Stochastic density functional theory (SDFT), also known as the Dean-Kawasaki equation~\cite{ dean1996langevin,kawasaki1994stochastic,te2020classical,Illien_2025}, is a powerful tool in statistical physics used for calculating dynamical properties and transport coefficients of a variety of condensed matter systems~\cite{demery2014generalized, kruger2018stresses} ranging from electrolytes ~\cite{demery2016conductivity,peraud2017fluctuation,avni2022conductivity,avni2022conductance,bonneau2023temporal,bonneau2024frequency,robin2024correlation} and polar liquids~\cite{varghese2025dynamic,dean2026dielectric}, to nonequilibrium binary mixtures~\cite{poncet2017universal,jardat2022diffusion,benois2023enhanced} and active matter~\cite{tailleur2008statistical, bertin2013mesoscopic,chavanis2008hamiltonian,gelimson2015collective}. 
At its core, SDFT is an exact reformulation  of many coupled Langevin equations, driven by Gaussian white noises, in terms of a single density field.
This itself is an approximation to a more fundamental description based on generalized Langevin equations with memory kernels \cite{zwanzig1973nonlinear,tuckerman2023statistical,zwanzig2001nonequilibrium,elber2019molecular}, which usually reduce to white noise when the environment's noise correlations decay on timescales much shorter than those of the process being described.

It turns out that even when the correlation time is vanishingly small, the noise may not be white~\cite{chun2018emergence}.
Indeed, the noise~$\zeta(t)$ in systems ranging from magnetized charged particles \cite{chun2018emergence,Balescu1997} to chiral active fluids \cite{Lowen2016,han2021fluctuating,caprini2023chiral} and 
open quantum systems \cite{Diosi1998} can be modeled by a complex Ornstein-Uhlenbeck process $\tau \,\dot{\zeta} = -(1+\ii \epsilon)\zeta + \xi$ (where $\xi$ is the standard complex Gaussian white noise) correlated over a timescale $\tau$ and endowed with a chirality characterized by the frequency $\omega\equiv\epsilon/\tau$ \cite{Arato1962,Chen2014,Metafune2002}.
We focus on a regime that we call \enquote{chiral overdamped limit}, where both the correlation time $\tau$ and the period $2\pi/\omega$ are negligible with respect to observation timescales while $\epsilon$ is fixed. In this limit,
the stochastic process $\zeta(t)$ is Gaussian, has a vanishing correlation time, but is not white, as it maintains a vestige of the underlying chirality.

\begin{figure}[t]
    \centering
    \includegraphics[width=\linewidth]{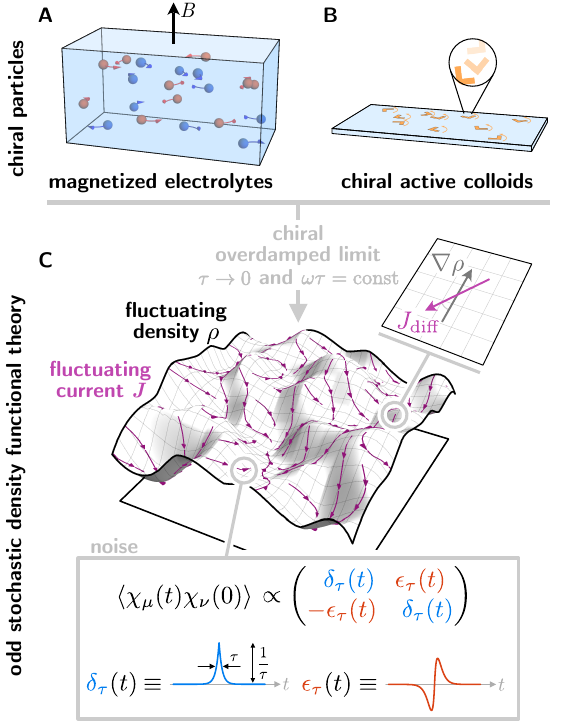}
    \vspace*{-0.7cm}
    \caption{
    \textbf{Odd stochastic density functional theory.}
    Chiral systems ranging from magnetized electrolytes (panel A) to chiral chemoactive colloids (panel B) can exhibit a chiral noise due to their coupling with the environment.
    These systems can be described by odd SDFT in the chiral overdamped limit where the correlation decay time $\tau$ of the noise vanishes with the same rate as the oscillation frequency of the chiral noise $\omega$ diverges.
    Odd SDFT (panel C) describes the evolution of a fluctuating density field $\rho$, which is redistributed by a fluctuating current $\bm J$. 
    The deterministic part of this current contains a diffusive part $\bm{J}_{\text{diff}}$ proportional to a rotated $\nabla \rho$ (plus drift currents), while the stochastic part, $\bm \chi$, exhibits correlations with a time-odd antisymmetric component.
    Indeed, the correlation matrix of the noise $\langle \chi_\mu(t) \chi_\nu(0)\rangle$ has a symmetric part proportional to the (nascent) Dirac distribution $\delta_\tau(t)$ (in blue) while the antisymmetric off-diagonal part is proportional to an antisymmetric function of time $\epsilon_\tau(t) = - \epsilon_\tau(-t)$ (in red).
    }
    \label{fig_chiral_sdft}
\end{figure}

In this Letter, we generalize SDFT to systems driven by such chiral noise. 
We dub this generalization \enquote{odd SDFT} because it exhibits correlation functions with an antisymmetric part that is odd under time-reversal symmetry (Fig.~\ref{fig_chiral_sdft}).
We demonstrate how odd SDFT provides a systematic route to compute odd (or Hall) response functions and transport coefficients that arise in chiral systems, such as diffusion, mobility or viscosity~\cite{fruchart2023odd}.
In particular, odd SDFT provides a framework for investigating the Hall effect in magnetized electrolytes, for which, to date, experimental measurements differ substantially from theoretical predictions~\cite{gerard1990hall,wolynes1980dynamics}.
Detailed derivations are presented in a companion paper, which primarily focuses on the
Hall conductance of magnetized electrolytes~\cite{avni2026hall}.

\medskip
{\it The chiral overdamped limit---}
Consider first a single charged particle subject to a  magnetic Lorentz force and in contact with a thermal bath at temperature $T$.
Here, the chiral overdamped limit arises in the small mass limit $m\to 0$, where the particle can be described by \cite{chun2018emergence}
\begin{equation} 
\label{ChunResult}
\dot{\boldsymbol{x}}=-\mathsf{M}\,\nabla U+\boldsymbol{\eta}(t)
\quad
\text{with}
\quad
\mathsf{M}\equiv
\gamma^{-1}
[\textsf{I}+\epsilon\,\bm{\varepsilon}_z]^{-1}
\end{equation}
where $\textsf{I}$ is the identity matrix, $\bm{\varepsilon}$ the Levi-Civita symbol \footnote{Here, $\bm{\varepsilon}$ is the 3D Levi-Civita symbol and $\bm{\varepsilon}_z$ is the matrix with elements $(\bm{\varepsilon}_z)_{ij}=\varepsilon_{zij}$. 
Equation \eqref{ChunResult} also holds in 2D upon replacing $\bm{\varepsilon}_z$ with the 2D Levi-Civita symbol.}, $\epsilon \equiv qB/\gamma$,  $\gamma$ the friction coefficient, $q$ the particle charge, $\bm{B}=B \hat{e}_z$ the magnetic field (taken along the $z$ axis for concreteness), and $U$ a potential (see Eq.~\eqref{Malpha} for $\mathsf{M}$ in matrix form). 
Crucially, due to the magnetic field, the noise $\bm \eta(t)$
is Gaussian but not white: it satisfies $\langle \bm \eta(t)\rangle=0$ and $\langle\bm{\eta}(t)\bm{\eta}^{\mathsf{T}}(t')\rangle
= \mathsf{C}(t-t')$ with \cite{chun2018emergence}
\begin{equation} \label{C_matrix}
\!\!\!
\mathsf{C}(t)
=
\lim_{\tau\to0}
\frac{k_B T}{\gamma\,\tau}
\ee^{-\frac{|t|}{\tau}}
\ee^{\frac{\epsilon t}{\tau} \bm{\varepsilon_z}}
=
k_B T [\mathsf{M}\delta_{+}(t)+\mathsf{M}^{\mathsf{T}}\delta_{-}(t)]
\end{equation}
where $\tau\equiv m/\gamma$ is the velocity relaxation timescale and $\omega_{\text{c}}\equiv\epsilon/\tau$ the cyclotron frequency, which simultaneously go to zero so that $\epsilon$ is fixed, and
where $\delta_{\pm}(u)$ are one-sided delta functions~\footnote{More precisely, $\delta_{+}(u)$ and $\delta_{-}(u)$ are defined to be zero if $u\neq 0$ and satisfy $\int\limits _{0}^{\infty}\delta_{+}\left(u\right){\rm d}u=\int\limits _{-\infty}^{0}\delta_{-}\left(u\right){\rm d}u=1
$ and $\int\limits _{0}^{\infty}\delta_{-}\left(u\right){\rm d}u=\int\limits _{-\infty}^{0}\delta_{+}\left(u\right){\rm d}u=0
$.}, while $\ee^{\alpha \bm{\varepsilon_z}}$ is the rotation matrix by an angle $\alpha$ around $z$.
The correlations of the nonwhite noise are related to the single-body mobility $\mathsf{M}$ via
\begin{equation}
    \label{FRT}
    \int_0^\infty \mathsf{C}(t) \dd t  = k_B T\,\mathsf{M}.
\end{equation}
This fluctuation-dissipation relation, yielding an asymmetric mobility $\mathsf{M} \neq \mathsf{M}^\TT$, could not occur with a Gaussian white noise (whose correlations are symmetric).

Beyond charged particles under magnetic field, Eq.~\eqref{ChunResult} has been proposed to model a broader class of systems called odd diffusive fluids~\cite{hargus2021odd,kalz2022collisions,kalz2026reversal,hargus2025odd,hargus2025flux,Abdoli2026}, characterized by an asymmetric friction matrix $\mathsf{M}^{-1}$, including skyrmions and passive objects in chiral baths.

\medskip
{\it Chiral active matter ---}
The chiral overdamped limit also arises in chiral active matter, a class of nonequilibrium systems made of active entities that are not mirror symmetric, ranging from chiral active colloids (Fig.~\ref{fig_chiral_sdft}B) to swimming bacteria~\cite{Liebchen2022,caprini2023chiral,kuroda2023microscopic}. 
Chiral active Brownian particles (ABP) are a paradigmatic representative of this class: these follow the dynamics
\begin{equation} 
\label{CABP}
\dot{\bm{x}}=-\frac{1}{\gamma}\nabla U+v_{0}\bm{n}[\theta(t)]
\quad
\text{and}
\quad
\dot{\theta}=\omega_{\text{R}}+\sqrt{\frac{2}{\tau_{\text{R}}}}\xi^{\text{R}}(t).
\end{equation}
where $\bm{n}[\theta]\equiv(\cos\theta,\sin\theta)$ is the orientation in the plane and $v_0$ is the self-propulsion speed, while $\omega_{\text{R}}$ encodes the chirality of the particles, $\tau_{\text{R}}$ is a persistence time, and $\xi^{\text{R}}$ is a standard Gaussian white noise (see Refs.~\cite{Kummel2013,Sevilla2016,Kuroda2025} for the 3D case).
Like in non-chiral active matter~\cite{martin2021statistical}, one can remove the non-Gaussian character of the resulting stochastic process $\bm{n}(t)$ by considering instead the chiral Ornstein-Uhlenbeck process $\dot{\bm{n}} =-\bm{n}/\tau + \omega \bm{n}\times\bm{z}
+\bm{\chi}/\sqrt{\tau}$ where $\bm{\chi}$ is a normalized Gaussian white noise, which results in a Gaussian colored noise $\bm{n}(t)$ with average $\langle \bm n(t)\rangle=0$ and correlations
\begin{equation} \label{nncorr}
    \langle\bm{n}(t)\bm{n}^{\mathsf{T}}(0)\rangle=\frac{1}{2}e^{-|t|/\tau_{\text{R}}}\left(\begin{array}{cc}
\cos\left(\omega_{\text{R}}t\right) & -\sin\left(\omega_{\text{R}}t\right)\\
\sin\left(\omega_{\text{R}}t\right) & \cos\left(\omega_{\text{R}}t\right)
\end{array}\right)
\end{equation}
identical to that of the chiral ABP (see Refs.~\cite{vanKampen1961,Kanazawa2012,Kanazawa2015,Fodor2018,Lucente2023} for non-Gaussian effects).

This stochastic process becomes white when $\tau_{\text{R}} \to 0$ at fixed $\omega_{\text{R}}$, but remains nonwhite with a vanishing correlation time in the chiral overdamped limit where both $\tau_{\text{R}} \to 0$ and $\omega_{\text{R}} \to \infty$ at fixed $\epsilon \equiv \omega_{\text{R}} \tau_{\text{R}}$. 
The model is then mapped onto
Eqs.~\eqref{ChunResult}--\eqref{C_matrix}
by replacing $\mathsf{M} \nabla U \to (1/\gamma) \nabla U$ and $k_{B}T \to {\gamma v_{0}^{2}\tau_{\rm R}}/{2}$ (so $v_0^2 \tau_{\text{R}}$
should also stay constant for the system to remain stochastic), but here the equivalent of the fluctuation-dissipation relation \eqref{FRT} does not hold because $\mathsf{M}$ is not the mobility.

\medskip
{\it The odd Dean-Kawasaki equation---}
The original derivation of SDFT relies on Ito calculus \cite{dean1996langevin}, which requires a white noise, so we cannot directly adapt it to Eq.~\eqref{ChunResult}, nor a many-body version thereof. 
Instead, we first go back to the underdamped stochastic dynamics
\begin{equation}
\label{underdamped_eom}
m\ddot{\boldsymbol{x}}_{i}=-{\partial U}/{\partial\boldsymbol{x}_{i}}-\mathsf{M}^{-1}\dot{\boldsymbol{x}}_{i}+\bm{\xi}_{i}(t).    
\end{equation}
for the positions $\bm{x}_i(t)$ of $N$ particles subject to a potential $U=\frac{1}{2}\sum_{i\neq j=1}^{N}V\left(|\boldsymbol{x}_{i}-\boldsymbol{x}_{j}|\right)+\sum_{i}V_{{\rm ext}}\left(\boldsymbol{x}_{i}\right)
$ made of pairwise interactions $V$ and an external potential $V_{\text{ext}}$, and subject to a vector Gaussian white noise $\bm \xi_{i}(t)$ with $\langle \bm \xi_{i}(t)\rangle$ and $\langle \xi_{i}^{\mu}(t)\xi_{j}^{\nu}(t')\rangle=2\gamma k_{\text{B}}T\delta_{ij}\delta_{\mu\nu}\delta\left(t-t'\right)$. 
Beyond charged particles in magnetic fields, Eq.~\eqref{underdamped_eom} can also arise from geometric (or Berry) forces, for instance in mesoscopic transport and in chemical reactions \cite{Bode2011,Bode2012,Dou2018,Bian2021,Tao2023}.
The dynamics \eqref{underdamped_eom} reduces to Eq.~\eqref{ChunResult} in the chiral overdamped (low mass) limit $m \to 0$ \cite{chun2018emergence}.
Then, generalizing the calculation of Ref.~\cite{nakamura2009derivation} for particles under a magnetic field, we  derive an underdamped fluctuating hydrodynamic equation for the number density $n(\bm x ,t)\equiv \sum_{i}^N\delta(\bm x-\bm x_i)$ and the momentum density $\boldsymbol{g}\left(\boldsymbol{x},t\right)\equiv \sum_{i}m\dot{\boldsymbol{x}}_{i}
$ (see Ref.~\cite{avni2026hall} for details), which yields the continuity equation $\partial_{t}n(\boldsymbol{x},t)=-\boldsymbol{\nabla}\cdot\boldsymbol{g}(\boldsymbol{x},t)/m
$ with
\begin{equation} \label{Eqg}
\partial_{t}g_{\mu}(\boldsymbol{x},t)=-\frac{(\mathsf{M}^{-1})_{\mu\nu}g_{\nu}}{m}-n\partial_{\mu}\frac{\delta U}{\delta n}-\partial_{\nu}\frac{g_{\mu}g_{\nu}}{m\,n}+\xi_{\mu}
\end{equation}
where $\mu,\nu$ denote Cartesian components, and where $\bm \xi$ is a Gaussian white noise vector field satisfying $\langle\bm \xi(\bm x,t)\rangle=0$ and $\langle\xi_{\mu}(\boldsymbol{x},t)\xi_{\nu}(\boldsymbol{x}',t')\rangle=2\gamma k_{B}Tn(\bm{x},t)\delta(\bm{x}-\bm{x}')\delta(t-t')\delta_{\mu\nu}
$.
Lastly, using the method of Ref.~\cite{nakamura2009derivation}, we adiabatically eliminate the momentum field $\bm{g}(\bm{x}, t)$ from Eq.~\eqref{Eqg} in the low-mass limit and get (see Ref.~\cite{avni2026hall} for details)
\begin{subequations}
\label{ODK}
\begin{align} \label{EOM1}
 \partial_{t}n  &=-\boldsymbol{\nabla}\cdot\boldsymbol{J}
\end{align}
with a current density
\begin{align}
 \label{EOM2}
 \boldsymbol{J} &= -k_{B}T
 \mathsf{M}
 \boldsymbol{\nabla}n-n
 \mathsf{M}
 \boldsymbol{\nabla}\frac{\delta U}{\delta n}+\bm{\chi}
\end{align}
\end{subequations}
where $\bm{\chi}(\bm x,t)$ is a Gaussian noise with $ \langle\bm\chi\left(\boldsymbol{x},t\right)\rangle = 0$ and 
\begin{equation} \label{noise3_corr}
\langle\chi_{\nu}(\boldsymbol{x},t)\chi_{\nu}(\bm{x}',t')\rangle=n(\bm{x},t)\delta(\bm{x}-\bm{x}')C_{\mu\nu}(t-t')
\end{equation}
with $\mathsf C$ given by Eq.~\eqref{C_matrix}. 
Equation~\eqref{ODK} implements the odd SDFT, and we refer to it as the ``odd Dean-Kawasaki equation''.
It reduces to the standard Dean-Kawasaki equation of Ref.~\cite{dean1996langevin} when the chirality parameter $\epsilon$ vanishes and to the dynamical density functional theory of Ref.~\cite{Abdoli2026} when noise vanishes.
For chiral active particles in the limit described above, it applies after the substitutions $\mathsf{M} \nabla \delta U/\delta n \to (1/\gamma) \nabla \delta U/\delta n$ and $k_{B}T \to {\gamma v_{0}^{2}\tau_{\rm R}}/{2}$.

\begin{figure}
\centering
{\includegraphics[width=0.5\textwidth,draft=false]{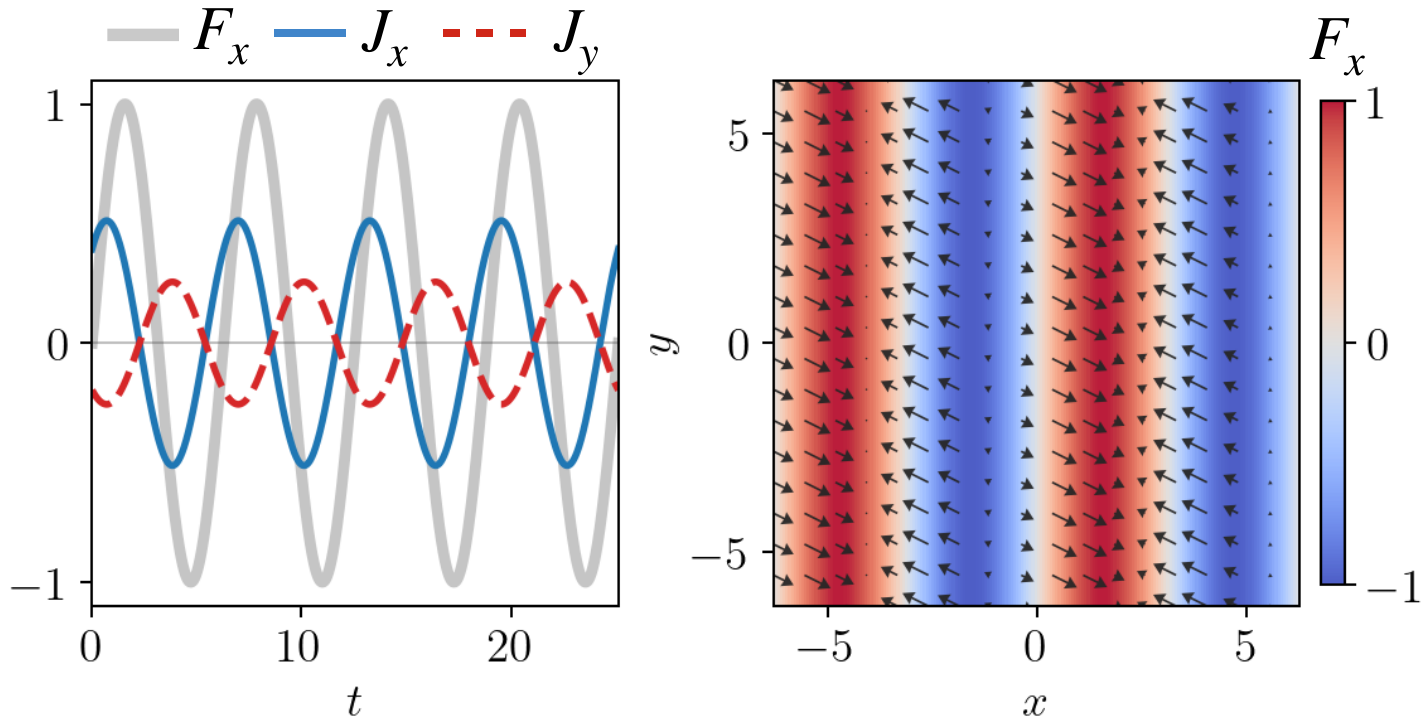}}
\caption{\textbf{Collective linear response to a spatiotemporal force from odd SDFT.} 
We apply a spatiotemporal force $\bm F=\sin(k_0x+\omega_0 t)\hat{x}$ and compute the resulting average current $\langle\bm{J}\rangle$.
Left: $F_x$ (thick gray), $J_x$ (full blue), and $J_y$ (dashed red) at the origin as a function of time. Right: $\langle \bm J(x,y)\rangle$ at $t=0$ (arrows), with colormap of $F_x(\bm x, t=0)$. The currents are calculated from inverse Fourier transform of Eqs.~\ref{def_Mkw} and~\ref{Mkw_result}. Parameters: $\varepsilon=0.5$, $k_BT={c}=\gamma=1$, $\varepsilon=0.5$,
$k_0=\omega_0=1$, and $\tilde{V}(k_0)=0.5$. 
\label{MobilityFig}}
\end{figure}

\medskip
{\it Collective mobility and Green-Kubo relation---} The odd Dean-Kawasaki equation can be used to determine the linear response of an ensemble of particles to an external force field $\bm{F}(\boldsymbol{x},t)=-\bm{\nabla} V_{\text{ext}}$. We characterize this response through the collective mobility tensor $\mathsf{L}$, defined in Fourier space \footnote{
We use the convention where $\tilde{f}\left(\boldsymbol{k},t\right)\equiv\int f(\boldsymbol{x},t)e^{-i \boldsymbol{k}\cdot\boldsymbol{x}}{\rm d}^{3}\boldsymbol{x}$
and
$\hat{f}\left(\boldsymbol{k},\omega\right)\equiv\int f\left(\boldsymbol{x},t\right)e^{-i\left(\boldsymbol{k}\cdot\boldsymbol{x}+\omega t\right)}{\rm d}^{3}\boldsymbol{x}\,{\rm d}t$.
} by
\begin{equation} \label{def_Mkw}
\langle\hat{J}_{\mu}(\boldsymbol{k},\omega)\rangle=\hat{L}_{\mu\nu}\hat{F}_{\nu}(\boldsymbol{k},\omega).
\end{equation}
Despite the broken time-reversal invariance inherent to chiral systems, and thanks to the fluctuation-dissipation relation~\eqref{FRT}, $\mathsf L$ satisfies the Green-Kubo relation (see proof in Ref.~\cite{avni2026hall})
\begin{equation} \label{GK_Mkw}
\!\!\!\!
\hat{L}_{\mu\nu}
(\boldsymbol{k},\omega)=\frac{1}{\mathcal{V}k_{B}T}\int_{0}^{\infty}
\!\!\!
\ee^{-\ii\omega t}\langle\tilde{{J}}_\mu\left(\boldsymbol{k},t\right){\tilde{J}_\nu}\left(-\boldsymbol{k},0\right)
\rangle_0{\rm d}t
\!\!
\end{equation}
where $\mathcal{V}$ is the system's volume, $\tilde{J}$ is the spatial Fourier transform of the current density in Eq.~\ref{EOM2}, and $\langle\cdot\rangle_0$ denotes ensemble average over the unperturbed system, i.e., at $\bm F=0$.
Using both a direct linear response calculation based on Eq.~\eqref{def_Mkw} and a Green-Kubo calculation based on Eq.~\eqref{GK_Mkw}, we compute the collective mobility at mean-field level~\cite{avni2026hall}. We find, using both methods, that
\begin{equation} \label{Mkw_result}
\hat{L}_{\mu\nu}(\boldsymbol{k},\omega)={c}\left[M_{\mu\nu}-\frac{M_{\mu\rho}M_{\sigma\nu}k_{\rho}k_{\sigma}}{M_{\varrho\varsigma}k_{\varrho}k_{\varsigma}+\frac{\ii\omega}{k_{\text{B}}T+{c}\tilde{V}(\bm{k})}}\right]
\end{equation}
where ${c}$ is the average bulk concentration. 
The right hand side of Eq.~\eqref{Mkw_result} is nonsymmetric due to the structure of $\mathsf{M}$ in Eq.~\eqref{ChunResult}, indicating an odd mobility tensor, and becomes symmetric for $\epsilon\to 0$ where $\mathsf{M}\to \gamma^{-1}\mathsf{I}$. In Fig.~\ref{MobilityFig}, we use Eq.~\ref{Mkw_result} to plot the current induced by a sinusoidal spatiotemporal force, $\bm F=\sin(k_0x+\omega_0t)\hat{x}$. It is shown that a transverse current develops under the applied force. When the external force is spatially homogeneous, the response is described by the $\bm k\to0$ limit of $\hat{\mathsf{L}}(\bm k, \omega)$ and is simply $\hat{\mathsf{L}}(\bm k \to 0, \omega)={c}\mathsf{M}$, which is notably independent of the interaction potential $V$. This can be understood by the fact that all particles drift together while their relative separations remain unchanged. 

\medskip
For chiral active Brownian particles, the substitution of $\mathsf{M}$ with $1/\gamma$ in front of the potential term in Eq.~\eqref{ODK} leads to a violation of the fluctuation-dissipation relation \eqref{FRT}, so the Green-Kubo relation~\eqref{GK_Mkw} does not hold (see Ref.~\cite{chun2021nonequilibrium} and references therein for nonequilibrium generalizations of Green-Kubo relations).
Nevertheless, the collective mobility of the chiral active particles can still be obtained by the direct linear response route (see Ref.~\cite{avni2026hall} for the full expression).

\medskip
{\it Electrolytes under magnetic fields---}
To describe electrolytes under magnetic fields, we extend the Odd Dean-Kawasaki equation to $N_{\rm s}$ ionic species with charges $q_{\alpha}$ and friction coefficients $\gamma_{\alpha}$, where $\alpha=1,...,N_{\rm s}$ denotes the species~\cite{avni2026hall}, yielding $\partial_{t}n_{\alpha}  =-\boldsymbol{\nabla}\cdot\boldsymbol{J}_{\alpha}$ and
 \begin{align} \label{ODKelectrolytes}
\boldsymbol{J}_{\alpha}(\bm{x},t) =  -k_{B}T\mathsf{M}_{\alpha}\boldsymbol{\nabla}n_{\alpha}+n_{\alpha}\mathsf{M}_{\alpha}q_{\alpha}\boldsymbol{E}+\boldsymbol{\chi}_{\alpha}\nonumber\\
  -n_{\alpha}\mathsf{M}_{\alpha}
 \sum_{\beta}
 \int\boldsymbol{\nabla}V_{\alpha\beta}(\bm{x}-\bm{x'})n_{\beta}(\bm{x}',t){\rm d}\bm{x}',
 \end{align}   
where the noise correlations are now given by $\langle\boldsymbol{\chi}_{\alpha}(\boldsymbol{x},t)\boldsymbol{\chi}_{\beta}^{\mathsf{T}}(\bm{x}',t')\rangle=n_{\alpha}(\boldsymbol{x},t)\delta_{\alpha\beta}\delta(\boldsymbol{x}-\boldsymbol{x}')\mathsf{C}_{\alpha}(t-t')
$
in which 
$\mathsf{C}_{\alpha}(t)  =k_B T[\mathsf{M}_{\alpha}\delta_{+}(t)+\mathsf{M}_{\alpha}^{\mathsf T}\delta_{-}(t)]$ 
and
\begin{equation}
\label{Malpha}
\mathsf{M_{\alpha}}  
\equiv\frac{1}{\gamma_{\alpha}}
\begin{pmatrix}
1 & -\frac{q_{\alpha}B}{\gamma_{\alpha}} & 0\\
\frac{q_{\alpha}B}{\gamma_{\alpha}} & 1 & 0\\
0 & 0 & 1
\end{pmatrix}^{-1}
\end{equation}
Here, $\bm E$ is an external electric field, $V_{\alpha\beta}(\boldsymbol{x})=q_{\alpha}q_{\beta}/4\pi\varepsilon_0\varepsilon_r|\boldsymbol{x}|
$ is the Coulomb interaction potential between two ions of charges $q_{\alpha}$ and $q_{\beta}$, $\varepsilon_{0}$ is the vacuum permittivity, and $\varepsilon_r$ the solvent's relative permittivity. 

In the linear-response regime, the electric conductivity tensor $\mathsf{K}$ relates the total charge current density
$\boldsymbol{J}_{c}\equiv\sum_{\alpha}q_{\alpha}\boldsymbol{J}_{\alpha}$
to the external electric field $\bm{E}$ through
\begin{equation} \label{cond_def}
\langle J_{c}^\mu\rangle=K_{\mu\nu}E_\nu
\end{equation}
where we assumed that the electric field $\bm E(\bm x, t) = E_0 \hat{n}$ is constant and uniform. 
Restricting ourselves to two-species symmetric electrolytes ($\alpha=\pm$ and $q_+=-q_-\equiv q$, with generally different $\gamma_+$ and $\gamma_-$), we calculate the conductivity from Eq.~\eqref{ODKelectrolytes} using three methods: direct linear response (via Eq.~\ref{cond_def}), the Green-Kubo relation
\begin{equation} \label{GKconductivity}
K_{\mu\nu}=\frac{1}{Vk_{B}T}\int\limits_{0}^{\infty}\!{\rm d}t\!\int\!{\rm d}\boldsymbol{x}\!\int\!{\rm d}\boldsymbol{x}'\!\langle J_{c}^{\mu}\left(\boldsymbol{x},t\right)J_{c}^{\nu}\left(\boldsymbol{x}',0\right)\rangle
\end{equation}
and a third hybrid fluctuation-dissipation relation derived from the perspective of first-order dynamics as in Refs.~\cite{Felderhof1983,Felderhof1987,Jardat1999,hoang2023frequency} (see Ref.~\cite{avni2026hall} for the derivation). 

In order to obtain the conductivity at small but finite concentrations, it is necessary to go beyond the mean-field level by performing the equivalent of a one-loop expansion in field theory. 
Our odd SDFT allows to perform such a calculation in a straightforward way.
We obtain the same result with all three methods: the conductivity tensor is given by \cite{avni2026hall}
\begin{equation}
\mathsf{K}
= \begin{pmatrix}
        \kappa & \kappa^{\text{H}}& {\color{gray} 0} \\
        - \kappa^{\text{H}} & \kappa & {\color{gray} 0} \\
        {\color{gray} 0} &{\color{gray} 0} & \kappa^{\parallel}
    \end{pmatrix}
\end{equation}

\begin{figure}
\centering
{\includegraphics[width=0.38\textwidth,draft=false]{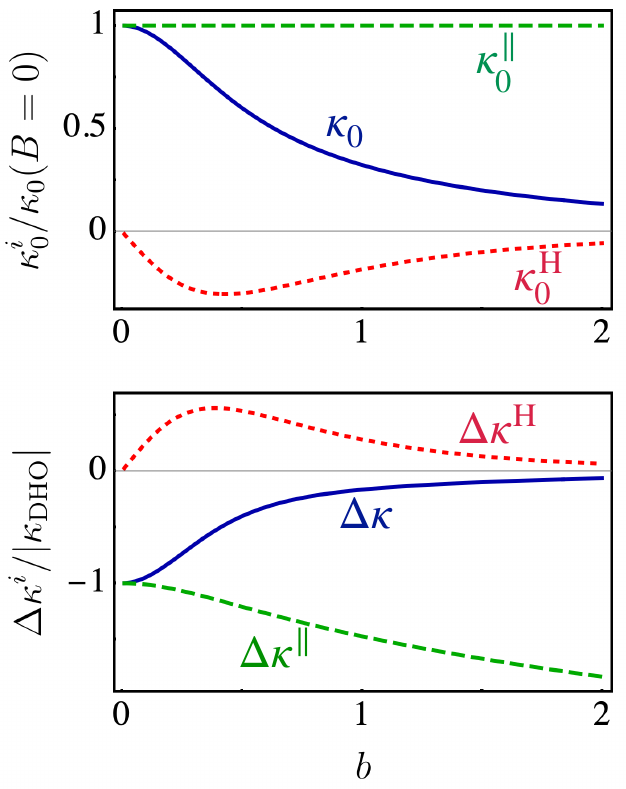}}
\caption{\textbf{Hall conductance of dilute electrolytes from odd SDFT.} The three independent components of the conductivity tensor, $\mathsf{K}_{xx}$, $\mathsf{K}_{xy}$, and $\mathsf{K}_{zz}$, shown as a function of $b=2q B/(\gamma_++\gamma_-)$, with fixed $2\gamma_+/(\gamma_++\gamma_-)=3/2$.
Top: the mean-field conductivity valid in the infinite dilution limit normalized by $\kappa_0(B=0)=q^2 {c} (\gamma_+^{-1}+\gamma_-^{-1})$. Bottom: lowest order correction at higher concentrations normalized by $|\Delta \kappa_{\rm DHO}|=\frac{\sqrt{2\pi}}{3}(2-\sqrt{2})\kappa_{0}(B=0)|z|^3\ell_{B}^{3/2}\sqrt{{c}}$. These results generalize the relaxation effect in the DHO theory to finite magnetic fields.
\label{HallCond}}
\end{figure}

\noindent with $\kappa\equiv \kappa_0 + \Delta \kappa$, $\kappa^{\text{H}}\equiv \kappa^{\text{H}}_0+\Delta\kappa^{\text{H}}$, and $\kappa^{\parallel}\equiv \kappa^{\parallel}_0 + \Delta \kappa^{\parallel}$. 
The terms $\kappa_0$, $\kappa^{\rm H}$, and $\kappa_0^{\parallel}$ are the mean-field conductivities, valid in the infinite dilution limit (${c}\to 0$), and given, to leading order in the magnetic field $B$, by
\begin{subequations}
\label{zeroth_order_conductivity}
\begin{align}
        \kappa_0 &= \kappa^{\parallel}_0=q^{2}{c}
        \big(\gamma_{+}^{-1} + \gamma_{-}^{-1}\big)+\mathcal{O}(B^2)
        \\
\kappa_{0}^{\text{H}}&=q^{3}{c}B\,\big(\gamma_{+}^{-2}-\gamma_{-}^{-2}\big)+\mathcal{O}(B^3).
\end{align}
\end{subequations}
The terms $\Delta\kappa$, $\Delta\kappa_{\rm H}$, and $\Delta\kappa$ are their respective corrections beyond mean field, scaling as ${c}^{3/2}$ and given by
\begin{subequations}
\label{first_order_conductivity}
\begin{align} \label{Deltakapaa1}
\!\! \Delta\kappa = \Delta\kappa^{\parallel} &=-\frac{\sqrt{2\pi}}{3}(2-\sqrt{2})
    \kappa_0
    |z|^3\ell_{B}^{3/2}
    \sqrt{{c}}+\mathcal{O}(B^2)
\\
\label{Deltakapaa2}
\Delta\kappa^{\text{H}} & =2qB(\gamma_{+}^{-1}-\gamma_{-}^{-1})\Delta\kappa+\mathcal{O}(B^{3})
\end{align}
\end{subequations}
where $\ell_{B} \equiv e^{2}/(4\pi\varepsilon_0\varepsilon_r k_{B}T)$ is the Bjerrum length, $z=q/e$ is the ionic valency and $e$ is the elementary charge. Note that the $c^{3/2}$ scaling becomes apparent after substituting $\kappa_0$ and $\Delta \kappa $ into Eqs.~\eqref{Deltakapaa1} and~\eqref{Deltakapaa2}, respectively. The full expressions as a function of $B$ (not restricted to linear order) are given in a companion paper~\cite{avni2026hall} and are plotted in Fig.~\ref{HallCond}, showing that both $|\kappa_0^{\rm H}|$ and $|\Delta\kappa^{\rm H}|$ peak at finite $B$ and approach zero in the $B\to\infty$ limit. Moreover, $\kappa_0$ and $\Delta \kappa$ deviate substantially from $\kappa_0^{\parallel}$ and $\Delta \kappa^{\parallel}$, respectively, beyond the linear regime, with the former diminishing at high $B$ while the latter remain finite. All three corrections reduce the conductivity, i.e., they have the opposite sign to the mean-field contribution (see~\cite{avni2026hall} for a more detailed discussion).
Our results reduce to that of the Debye-H{\"u}ckel-Onsager (DHO) theory~\cite{onsager1927report,onsager2002irreversible} at zero magnetic field ($B=0$), and also match the results of a cluster expansion performed by H. Friedman in 1965~\cite{friedman1965calculation} that predicts the first order correction in $B$ (see Ref.~\cite{avni2026hall} for details).

Our results ignore two important effects related to the solvent: 
(i) fluid advection, which can be accounted for by coupling the odd Dean-Kawasaki equation with fluctuating hydrodynamics~\cite{peraud2017fluctuation,avni2022conductance}, and leads to so-called electrophoretic terms in the DHO theory (neglected here), and 
(ii) the dielectric response of the solvent, which may be taken into account through an appropriate model for the moving ion-induced polarization of the solvent in a magnetic field~\cite{hubbard1981electrohydrodynamic,sung1987microscopic}.
These effects are believed to produce an additional drag on the ion that can change the Hall conductivity even in the infinite-dilution limit~\cite{wolynes1980dynamics,sung1987microscopic,Friedman1968,Harris1970,Harris1972,Khanh1972,hubbard1981electrohydrodynamic,Kroh1988,Chechetkin1989,kroh1990hall}.
Such an effect has indeed been reported in experimental measurements of the Hall effect in electrolytes~\cite{Mergault1964,LaforgueKantzer1964,LaforgueKantzer1965,Wendhausen1968,Hanebeck1974,Gerard1990,Frank1972,Abbes1982,Bellissent1971,meton1976hall,Meton1976b,Abbes1980,Hanebeck1974}, but are not satisfactorily captured by theoretical predictions, which in particular do not account for the observed difference in behavior between cations and anions with the same ionic radius~\cite{gerard1990hall,wolynes1980dynamics}.
This mismatch points towards the need of taking into account the dielectric response of the solvent more finely, which our stochastic field theoretical approach could enable by combining the effect of magnetic fields using the odd SDFT described in this Letter with modern approaches to describe the structured dielectric response of solvents \cite{Kornyshev1986,Belaya1986,Levy2012,Berthoumieux2019,Berthoumieux2021,Illien2024,Blossey2022,Blossey2022b,Berthoumieux2024,adar2018dielectric}.

\medskip
{\it Conclusions---} We have introduced an odd stochastic density functional theory to describe the chiral overdamped limit of systems ranging from magnetized electrolytes to active fluids, where the memoryless noise keeps a trace of the chirality.
This framework allows us to systematically compute the asymmetric response of these chiral systems beyond mean-field, including odd mobility and Hall conductivity.

\bigskip\bigskip

\begin{acknowledgments}
Y.A. dedicates this work to the memory of Rudolf Podgornik, a generous mentor and a true gentleman. V.V. acknowledges partial support from the Army Research Office under grant W911NF-22-2-0109 and W911NF-23-1-
0212. M.F. and V.V acknowledge partial support from
the France Chicago center through a FACCTS grant.
This research was partly supported from the National
Science Foundation through the Center for Living Systems (grant no. 2317138), the Simons Foundation and
the Chan Zuckerberg Foundation.
\end{acknowledgments}

\bibliography{bibliography}

@article{dean1996langevin,
  title={Langevin equation for the density of a system of interacting Langevin processes},
  author={Dean, David S},
  journal={Journal of Physics A: Mathematical and General},
  volume={29},
  number={24},
  pages={L613},
  year={1996},
  publisher={IOP Publishing},
  doi={10.1088/0305-4470/29/24/001}
}

@article{demery2016conductivity,
  doi = {10.1088/1742-5468/2016/02/023106},
  title={The conductivity of strong electrolytes from stochastic density functional theory},
  author={D{\'e}mery, Vincent and Dean, David S},
  journal={Journal of Statistical Mechanics: Theory and Experiment},
  volume={2016},
  number={2},
  pages={023106},
  year={2016},
  publisher={IOP Publishing}
}

@book{zwanzig2001nonequilibrium,
  title={Nonequilibrium statistical mechanics},
  author={Zwanzig, Robert},
  year={2001},
  publisher={Oxford university press}
}

@book{tuckerman2023statistical,
  doi={10.1093/oso/9780198825562.001.0001},
  isbn={9780198825562},
  title={Statistical mechanics: theory and molecular simulation},
  author={Tuckerman, Mark E},
  year={2023},
  publisher={Oxford university press}
}

@article{Illien_2025,
doi = {10.1088/1361-6633/adee2e},
url = {https://doi.org/10.1088/1361-6633/adee2e},
year = {2025},
month = {aug},
publisher = {IOP Publishing},
volume = {88},
number = {8},
pages = {086601},
author = {Illien, Pierre},
title = {The Dean–Kawasaki equation and stochastic density functional theory},
journal = {Reports on Progress in Physics}
}

@article{kawasaki1994stochastic,
  title={Stochastic model of slow dynamics in supercooled liquids and dense colloidal suspensions},
  author={Kawasaki, Kyozi},
  journal={Physica A: Statistical Mechanics and its Applications},
  volume={208},
  number={1},
  pages={35--64},
  year={1994},
  publisher={Elsevier}
}

@article{peraud2017fluctuation,
doi = {10.1073/pnas.1714464114},
 title = {Fluctuation-enhanced electric conductivity in electrolyte solutions},
 author = {Péraud, Jean-Philippe and Nonaka, Andrew J. and Bell, John B. and Donev, Aleksandar and Garcia, Alejandro L.},
 journal = {Proceedings of the National Academy of Sciences},
 volume = {114},
 number = {41},
 pages = {10829–10833},
 year = {2017},
 month = {sept},
 issn = {1091-6490},
 publisher = {National Academy of Sciences}
}

@article{avni2022conductivity,
  doi = {10.1103/physrevlett.128.098002},
  title={Conductivity of concentrated electrolytes},
  author={Avni, Yael and Adar, Ram M and Andelman, David and Orland, Henri},
  journal={Physical Review Letters},
  volume={128},
  number={9},
  pages={098002},
  year={2022},
  publisher={APS}
}

@article{han2021fluctuating,
  doi = {10.1038/s41567-021-01360-7},
  title={Fluctuating hydrodynamics of chiral active fluids},
  author={Han, Ming and Fruchart, Michel and Scheibner, Colin and Vaikuntanathan, Suriyanarayanan and De Pablo, Juan J and Vitelli, Vincenzo},
  journal={Nature Physics},
  volume={17},
  number={11},
  pages={1260--1269},
  year={2021},
  publisher={Nature Publishing Group UK London}
}

@article{robin2024correlation,
  doi = {10.1063/5.0188215},
  title={Correlation-induced viscous dissipation in concentrated electrolytes},
  author={Robin, Paul},
  journal={The Journal of Chemical Physics},
  volume={160},
  number={6},
  year={2024},
  pages={064503},
  publisher={AIP Publishing}
}

@article{te2020classical,
  doi = {10.1080/00018732.2020.1854965},
  title={Classical dynamical density functional theory: from fundamentals to applications},
  author={te Vrugt, Michael and L{\"o}wen, Hartmut and Wittkowski, Raphael},
  journal={Advances in Physics},
  volume={69},
  number={2},
  pages={121--247},
  year={2020},
  publisher={Taylor \& Francis}
}

@article{benois2023enhanced,
  doi = {10.1103/physreve.108.054606},
  title={Enhanced diffusion of tracer particles in nonreciprocal mixtures},
  author={Benois, Anthony and Jardat, Marie and Dahirel, Vincent and D{\'e}mery, Vincent and Agudo-Canalejo, Jaime and Golestanian, Ramin and Illien, Pierre},
  journal={Physical Review E},
  volume={108},
  number={5},
  pages={054606},
  year={2023},
  publisher={APS}
}

@article{chun2018emergence,
  doi = {10.1103/physreve.97.032117},
  title={Emergence of nonwhite noise in Langevin dynamics with magnetic Lorentz force},
  author={Chun, Hyun-Myung and Durang, Xavier and Noh, Jae Dong},
  journal={Physical Review E},
  volume={97},
  number={3},
  pages={032117},
  year={2018},
  publisher={APS}
}

@article{chun2021nonequilibrium,
  doi = {10.1103/physrevresearch.3.043172},
  title={Nonequilibrium Green-Kubo relations for hydrodynamic transport from an equilibrium-like fluctuation-response equality},
  author={Chun, Hyun-Myung and Gao, Qi and Horowitz, Jordan M},
  journal={Physical Review Research},
  volume={3},
  number={4},
  pages={043172},
  year={2021},
  publisher={APS}
}

@article{nakamura2009derivation,
  doi = {10.1088/1751-8113/42/6/065001},
  title={Derivation of the nonlinear fluctuating hydrodynamic equation from the underdamped Langevin equation},
  author={Nakamura, Takenobu and Yoshimori, Akira},
  journal={Journal of Physics A: Mathematical and Theoretical},
  volume={42},
  number={6},
  pages={065001},
  year={2009},
  publisher={IOP Publishing}
}

@article{hoang2023frequency,
 doi = {10.1063/5.0139258},
 title = {Frequency and field-dependent response of confined electrolytes from Brownian dynamics simulations},
 author = {Hoang Ngoc Minh, Thê and Stoltz, Gabriel and Rotenberg, Benjamin},
 journal = {The Journal of Chemical Physics},
 volume = {158},
 number = {10},
 year = {2023},
 month = {mar},
 pages = {104103},
 issn = {1089-7690},
 publisher = {AIP Publishing}
}

@misc{dean2026dielectric,
 eprint = {2604.00262},
 title = {Dielectric response as a source of viscosity in polar liquids},
 author = {David S. Dean and Haim Diamant},
 year = {2026},
 eprinttype = {arXiv},
 archivePrefix = {arXiv}
}

@article{kruger2018stresses,
  doi = {10.1063/1.5019424},
  title={Stresses in non-equilibrium fluids: Exact formulation and coarse-grained theory},
  author={Kr{\"u}ger, Matthias and Solon, Alexandre and D{\'e}mery, Vincent and Rohwer, Christian M and Dean, David S},
  journal={The Journal of chemical physics},
  volume={148},
  number={8},
  year={2018},
  pages={084503},
  publisher={AIP Publishing}
}

@article{avni2022conductance,
  doi = {10.1063/5.0111645},
  title={Conductance of concentrated electrolytes: Multivalency and the Wien effect},
  author={Avni, Yael and Andelman, David and Orland, Henri},
  journal={The Journal of Chemical Physics},
  volume={157},
  number={15},
  year={2022},
  pages={154502},
  publisher={AIP Publishing}
}

@article{bonneau2024frequency,
  doi = {10.1063/5.0236073},
  title={Frequency-dependent conductivity of concentrated electrolytes: A stochastic density functional theory},
  author={Bonneau, Haggai and Avni, Yael and Andelman, David and Orland, Henri},
  journal={The Journal of Chemical Physics},
  volume={161},
  number={24},
  year={2024},
  pages={244501},
  publisher={AIP Publishing}
}

@article{bonneau2023temporal,
  doi = {10.1088/1742-5468/acdced},
  title={Temporal response of the conductivity of electrolytes},
  author={Bonneau, Haggai and D{\'e}mery, Vincent and Rapha{\"e}l, Elie},
  journal={Journal of Statistical Mechanics: Theory and Experiment},
  volume={2023},
  number={7},
  pages={073205},
  year={2023},
  publisher={IOP Publishing}
}

@article{fruchart2023odd,
  doi = {10.1146/annurev-conmatphys-040821-125506},
  title={Odd viscosity and odd elasticity},
  author={Fruchart, Michel and Scheibner, Colin and Vitelli, Vincenzo},
  journal={Annual Review of Condensed Matter Physics},
  volume={14},
  number={1},
  pages={471--510},
  year={2023},
  publisher={Annual Reviews}
}

@article{onsager1927report,
  doi = {10.1039/tf9272300341},
  title={Report on a revision of the conductivity theory},
  author={Onsager, Lars},
  journal={Transactions of the Faraday Society},
  volume={23},
  pages={341--349},
  year={1927},
  publisher={Royal Society of Chemistry}
}

@article{onsager2002irreversible,
  doi={10.1021/j150341a001},
  title={Irreversible processes in electrolytes. Diffusion, conductance and viscous flow in arbitrary mixtures of strong electrolytes},
  author={Onsager, Lars and Fuoss, Raymond Matthew},
  journal={The Journal of Physical Chemistry},
  volume={36},
  number={11},
  pages={2689--2778},
  year={2002},
  publisher={ACS Publications}
}

@article{caprini2023chiral,
  title={Chiral active matter in external potentials},
  author={Caprini, Lorenzo and L{\"o}wen, Hartmut and Marconi, Umberto Marini Bettolo},
  journal={Soft Matter},
  volume={19},
  number={33},
  pages={6234--6246},
  year={2023},
  publisher={Royal Society of Chemistry},
  doi={10.1039/d3sm00793f}
}

@article{demery2014generalized,
  doi = {10.1088/1367-2630/16/5/053032},
  title={Generalized Langevin equations for a driven tracer in dense soft colloids: construction and applications},
  author={D{\'e}mery, Vincent and B{\'e}nichou, Olivier and Jacquin, Hugo},
  journal={New Journal of Physics},
  volume={16},
  number={5},
  pages={053032},
  year={2014},
  publisher={IOP Publishing}
}

@article{friedman1965calculation,
 doi = {10.1021/j100892a023},
 title = {Calculation of the Hall Effect in Ionic Solutions},
 author = {Friedman, Harold L.},
 journal = {The Journal of Physical Chemistry},
 volume = {69},
 number = {8},
 pages = {2617–2628},
 year = {1965},
 month = {aug},
 issn = {1541-5740},
 publisher = {American Chemical Society (ACS)}
}

@article{meton1976hall,
 doi = {10.1016/0009-2614(76)80733-6},
 title = {Hall effect in dilute electrolytes},
 author = {Meton, Maurice and Gerard, Paul},
 journal = {Chemical Physics Letters},
 volume = {44},
 number = {3},
 pages = {582–585},
 year = {1976},
 month = {dec},
 issn = {0009-2614},
 publisher = {Elsevier}
}

@article{Abdoli2026,
 doi = {10.1021/acs.jpcb.6c00652},
 title = {Dynamical Density Functional Theory for Dense Odd-Diffusive Fluids},
 author = {Abdoli, Iman and Wittmann, René and Löwen, Hartmut},
 journal = {The Journal of Physical Chemistry B},
 volume = {130},
 number = {17},
 pages = {4672–4682},
 year = {2026},
 month = {apr},
 issn = {1520-5207},
 publisher = {American Chemical Society (ACS)}
}

@article{jardat2022diffusion,
 doi = {10.1103/physreve.106.064608},
 title = {Diffusion of a tracer in a dense mixture of soft particles connected to different thermostats},
 author = {Jardat, Marie and Dahirel, Vincent and Illien, Pierre},
 journal = {Physical Review E},
 volume = {106},
 number = {6},
 pages = {064608},
 year = {2022},
 month = {dec},
 issn = {2470-0053},
 publisher = {American Physical Society (APS)}
}

@article{poncet2017universal,
  doi = {10.1103/physrevlett.118.118002},
 title = {Universal Long Ranged Correlations in Driven Binary Mixtures},
 author = {Poncet, Alexis and Bénichou, Olivier and Démery, Vincent and Oshanin, Gleb},
 journal = {Physical Review Letters},
 volume = {118},
 number = {11},
 pages = {118002},
 year = {2017},
 month = {mar},
 issn = {1079-7114},
 publisher = {American Physical Society (APS)}
}

@article{sung1987microscopic,
 doi = {10.1063/1.453559},
 title = {The microscopic theory of the Hall effect in ionic solutions. I. General formulation and the Brownian ion-continuum solvent limit},
 author = {Sung, Wokyung and Friedman, Harold L.},
 journal = {The Journal of Chemical Physics},
 volume = {87},
 number = {1},
 pages = {643–653},
 year = {1987},
 month = {july},
 issn = {1089-7690},
 publisher = {AIP Publishing}
}

@article{hubbard1981electrohydrodynamic,
 doi = {10.1063/1.442400},
 title = {An electrohydrodynamic contribution to the Hall effect in electrolyte solutions},
 author = {Hubbard, J. B. and Wolynes, P. G.},
 journal = {The Journal of Chemical Physics},
 volume = {75},
 number = {6},
 pages = {3051–3054},
 year = {1981},
 month = {sept},
 issn = {1089-7690},
 publisher = {AIP Publishing}
}

@article{kroh1990hall,
 doi = {10.1080/00268979000100881},
 title = {The Hall effect in dilute ionic solutions},
 author = {Kroh, H.J. and Felderhof, B.U.},
 journal = {Molecular Physics},
 volume = {70},
 number = {1},
 pages = {119–128},
 year = {1990},
 month = {may},
 issn = {1362-3028},
 publisher = {Informa UK Limited}
}

@article{gerard1990hall,
  title={Hall effect in aqueous acid solutions at different concentrations},
  author={Gerard, P and Gerard, R and Meton, M and Picard, EJ},
  journal={Journal of the Electrochemical Society},
  volume={137},
  number={12},
  pages={3873--3875},
  year={1990},
  publisher={The Electrochemical Society, Inc.},
  doi = {10.1149/1.2086318},
}

@article{wolynes1980dynamics,
 doi = {10.1146/annurev.pc.31.100180.002021},
 title = {Dynamics of Electrolyte Solutions},
 author = {Wolynes, P G},
 journal = {Annual Review of Physical Chemistry},
 volume = {31},
 number = {1},
 pages = {345–376},
 year = {1980},
 month = {oct},
 issn = {1545-1593},
 publisher = {Annual Reviews}
}

@article{Mergault1964,
title={Mesure d'effets magnétoélectriques dans des solutions aqueuses d'électrolytes},
author={Pierre Mergault and Josette Pagès-Nelson},
journal={Comptes Rendus de l'Académie des Sciences},
volume={259},
pages={4656--4659},
year=1964,
url={https://gallica.bnf.fr/ark:/12148/bpt6k4015m/f1786.item}
}

@article{Gerard1990,
 doi = {10.1149/1.2086318},
 title = {Hall Effect in Aqueous Acid Solutions at Different Concentrations},
 author = {Gérard, P. and Gérard, R. and Meton, M. and Picard, E. J.},
 journal = {Journal of The Electrochemical Society},
 volume = {137},
 number = {12},
 pages = {3873–3875},
 year = {1990},
 month = {dec},
 issn = {1945-7111},
 publisher = {The Electrochemical Society}
}

@article{Frank1972,
 doi = {10.1080/00319107208084099},
 title = {Hall voltage in electrolyte solutions},
 author = {Frank, Ronald L. and Hoffman, Joseph G.},
 journal = {Physics and Chemistry of Liquids},
 volume = {3},
 number = {4},
 pages = {191–204},
 year = {1972},
 month = {jan},
 issn = {1029-0451},
 publisher = {Informa UK Limited}
}

@article{Abbes1982,
 doi = {10.1051/jphyslet:01982004305012700},
 title = {Effet Hall de chlorures et d’iodures en solutions diluées dans le méthanol et dans l’eau},
 author = {Abbes, M. and Gérard, R. and Gérard, P. and Meton, M. and Picard, E.J.},
 journal = {Journal de Physique Lettres},
 volume = {43},
 number = {5},
 pages = {127–132},
 year = {1982},
 issn = {0302-072X},
 publisher = {EDP Sciences}
}

@article{Bellissent1971,
 doi = {10.1149/1.2407872},
 title = {Eperimental Measurements of the Hall Effect in Very Dilute Solutions},
 author = {Bellissent, Marie-Claire and Gerard, Paul and Longevialle, Christian and Meton, Maurice and Pich, Michel and Morand, Geneviève},
 journal = {Journal of The Electrochemical Society},
 volume = {118},
 number = {12},
 pages = {1944},
 year = {1971},
 issn = {0013-4651},
 publisher = {The Electrochemical Society}
}

@article{Meton1976b,
 doi = {10.1051/jphyslet:019760037010024700},
 title = {Effet hall dans les solutions électrolytiques diluées, solvatation et dynamique des ions},
 author = {Meton, M. and Gerard, P. and Picard, E.J.},
 journal = {Journal de Physique Lettres},
 volume = {37},
 number = {10},
 pages = {247–250},
 year = {1976},
 issn = {0302-072X},
 publisher = {EDP Sciences}
}

@article{tailleur2008statistical,
doi = {10.1103/physrevlett.100.218103},
  title={Statistical mechanics of interacting run-and-tumble bacteria},
  author={Tailleur, Julien and Cates, Michael E},
  journal={Physical review letters},
  volume={100},
  number={21},
  pages={218103},
  year={2008},
  publisher={APS}
}

@article{bertin2013mesoscopic,
  doi = {10.1088/1367-2630/15/8/085032},
  title={Mesoscopic theory for fluctuating active nematics},
  author={Bertin, Eric and Chat{\'e}, Hugues and Ginelli, Francesco and Mishra, Shradha and Peshkov, Anton and Ramaswamy, Sriram},
  journal={New journal of physics},
  volume={15},
  number={8},
  pages={085032},
  year={2013},
  publisher={IOP Publishing}
}

@article{kuroda2023microscopic,
  doi = {10.1088/1742-5468/ad0639},
  title={Microscopic theory for hyperuniformity in two-dimensional chiral active fluid},
  author={Kuroda, Yuta and Miyazaki, Kunimasa},
  journal={Journal of Statistical Mechanics: Theory and Experiment},
  volume={2023},
  number={10},
  pages={103203},
  year={2023},
  publisher={IOP Publishing}
}

@article{gelimson2015collective,
  doi = {10.1103/physrevlett.114.028101},
  title={Collective dynamics of dividing chemotactic cells},
  author={Gelimson, Anatolij and Golestanian, Ramin},
  journal={Physical review letters},
  volume={114},
  number={2},
  pages={028101},
  year={2015},
  publisher={APS}
}

@article{chavanis2008hamiltonian,
  doi = {10.1016/j.physa.2008.06.016},
  title={Hamiltonian and Brownian systems with long-range interactions: V. Stochastic kinetic equations and theory of fluctuations},
  author={Chavanis, Pierre-Henri},
  journal={Physica A: Statistical Mechanics and its Applications},
  volume={387},
  number={23},
  pages={5716--5740},
  year={2008},
  publisher={Elsevier}
}

@article{varghese2025dynamic,
  doi = {10.1063/5.0292306},
  title={Dynamic correlations in a polar fluid: confronting stochastic density functional theory to simulations},
  author={Varghese, Sleeba and Illien, Pierre and Rotenberg, Benjamin},
  journal={The Journal of Chemical Physics},
  volume={163},
  number={12},
  year={2025},
  pages={124107},
  publisher={AIP Publishing}
}

@article{Abbes1980,
 doi = {10.1051/jphyslet:019800041023057500},
 title = {Effet Hall dans des solutions électrolytiques de concentration variable},
 author = {Abbes, M. and Gérard, R. and Gérard, P. and Meton, M. and Picard, E.J.},
 journal = {Journal de Physique Lettres},
 volume = {41},
 number = {23},
 pages = {575–580},
 year = {1980},
 issn = {0302-072X},
 publisher = {EDP Sciences}
}

@article{LaforgueKantzer1965,
 doi = {10.1016/0013-4686(65)87038-4},
 title = {Effet magnetoelectrique des solutions d’acides mineraux},
 author = {Laforgue-Kantzer, D.},
 journal = {Electrochimica Acta},
 volume = {10},
 number = {6},
 pages = {585–603},
 year = {1965},
 month = {june},
 issn = {0013-4686},
 publisher = {Elsevier BV}
}

@article{kalz2022collisions,
 doi = {10.1103/physrevlett.129.090601},
  title={Collisions enhance self-diffusion in odd-diffusive systems},
  author={Kalz, Erik and Vuijk, Hidde Derk and Abdoli, Iman and Sommer, Jens-Uwe and L{\"o}wen, Hartmut and Sharma, Abhinav},
  journal={Physical Review Letters},
  volume={129},
  number={9},
  pages={090601},
  year={2022},
  publisher={APS}
}

@article{kalz2026reversal,
  doi = {10.1103/k8vq-dw2w},
  title={Reversal of tracer advection and Hall drift in an interacting chiral fluid},
  author={Kalz, Erik and Ravichandir, Shashank and Birkenmeier, Johannes and Metzler, Ralf and Sharma, Abhinav},
  journal={Physical Review E},
  volume={113},
  number={4},
  pages={L042104},
  year={2026},
  publisher={APS}
}

@article{hargus2021odd,
 doi = {10.1103/physrevlett.127.178001},
 title = {Odd Diffusivity of Chiral Random Motion},
 author = {Hargus, Cory and Epstein, Jeffrey M. and Mandadapu, Kranthi K.},
 journal = {Physical Review Letters},
 volume = {127},
 number = {17},
 pages = {178001},
 year = {2021},
 month = {oct},
 issn = {1079-7114},
 publisher = {American Physical Society (APS)}
}

@article{hargus2025odd,
 doi = {10.1103/pdpf-sd9q},
  title={Odd dynamics of passive objects in a chiral active bath},
  author={Hargus, Cory and Ghimenti, Federico and Tailleur, Julien and van Wijland, Fr{\'e}d{\'e}ric},
  journal={Physical Review Letters},
  volume={135},
  number={16},
  pages={167102},
  year={2025},
  publisher={APS}
}

@article{Hanebeck1974,
 doi = {10.1002/bbpc.19740780405},
 title = {Über den Hall‐Effekt an einigen Elektrolytlösungen und Membranen},
 author = {Hanebeck, N. and Schmid, G.},
 journal = {Berichte der Bunsengesellschaft für physikalische Chemie},
 volume = {78},
 number = {4},
 pages = {325–331},
 year = {1974},
 month = {apr},
 issn = {0005-9021},
 publisher = {Wiley}
}

@article{Wendhausen1968,
 doi = {10.1524/zpch.1968.58.5_6.325},
 title = {Messungen des Hall-Effektes in wäßrigen Elektrolytlösungen},
 author = {Wendhausen, Henning},
 journal = {Zeitschrift für Physikalische Chemie},
 volume = {58},
 number = {5_6},
 pages = {325–326},
 year = {1968},
 month = {apr},
 issn = {0942-9352},
 publisher = {Walter de Gruyter GmbH}
}

@article{Blossey2022,
 doi = {10.1103/physrevresearch.4.023033},
 title = {Field theory of structured liquid dielectrics},
 author = {Blossey, Ralf and Podgornik, Rudolf},
 journal = {Physical Review Research},
 volume = {4},
 number = {2},
 pages = {023033},
 year = {2022},
 month = {apr},
 issn = {2643-1564},
 publisher = {American Physical Society (APS)}
}

@article{Berthoumieux2021,
 doi = {10.1063/5.0056430},
 title = {Dipolar Poisson models in a dual view},
 author = {Berthoumieux, Hélène and Monet, Geoffrey and Blossey, Ralf},
 journal = {The Journal of Chemical Physics},
 volume = {155},
 number = {2},
 year = {2021},
 month = {july},
 issn = {1089-7690},
 pages={024112},
 publisher = {AIP Publishing}
}

@article{Berthoumieux2019,
 doi = {10.1063/1.5080183},
 title = {Dielectric response in the vicinity of an ion: A nonlocal and nonlinear model of the dielectric properties of water},
 author = {Berthoumieux, H. and Paillusson, F.},
 journal = {The Journal of Chemical Physics},
 volume = {150},
 number = {9},
 year = {2019},
 month = {mar},
 pages = {094507},
 issn = {1089-7690},
 publisher = {AIP Publishing}
}

@article{Levy2012,
 doi = {10.1103/physrevlett.108.227801},
 title = {Dielectric Constant of Ionic Solutions: A Field-Theory Approach},
 author = {Levy, Amir and Andelman, David and Orland, Henri},
 journal = {Physical Review Letters},
 volume = {108},
 number = {22},
 pages = {227801},
 year = {2012},
 month = {may},
 issn = {1079-7114},
 publisher = {American Physical Society (APS)}
}

@article{Kornyshev1986,
 doi = {10.1016/0022-0728(86)80509-5},
 title = {On the non-local electrostatic theory of hydration force},
 author = {Kornyshev, A.A.},
 journal = {Journal of Electroanalytical Chemistry and Interfacial Electrochemistry},
 volume = {204},
 number = {1-2},
 pages = {79–84},
 year = {1986},
 month = {june},
 issn = {0022-0728},
 publisher = {Elsevier BV}
}

@article{Belaya1986,
 doi = {10.1016/s0009-2614(86)80099-9},
 title = {Hydration forces as a result of non-local water polarizability},
 author = {Belaya, M.L. and Feigel’man, M.V. and Levadny, V.G.},
 journal = {Chemical Physics Letters},
 volume = {126},
 number = {3-4},
 pages = {361–364},
 year = {1986},
 month = {may},
 issn = {0009-2614},
 publisher = {Elsevier BV}
}

@article{Illien2024,
 doi = {10.1103/physrevlett.133.268002},
 title = {Stochastic Density Functional Theory for Ions in a Polar Solvent},
 author = {Illien, Pierre and Carof, Antoine and Rotenberg, Benjamin},
 journal = {Physical Review Letters},
 volume = {133},
 number = {26},
 pages = {268002},
 year = {2024},
 month = {dec},
 issn = {1079-7114},
 publisher = {American Physical Society (APS)}
}

@article{Blossey2022b,
 doi = {10.1209/0295-5075/ac7d0a},
 title = {Continuum theories of structured dielectrics},
 author = {Blossey, Ralf and Podgornik, Rudolf},
 journal = {Europhysics Letters},
 volume = {139},
 number = {2},
 pages = {27002},
 year = {2022},
 month = {july},
 issn = {1286-4854},
 publisher = {IOP Publishing}
}

@article{Berthoumieux2024,
 doi = {10.1063/5.0226773},
 title = {Nonlinear conductivity of aqueous electrolytes: Beyond the first Wien effect},
 author = {Berthoumieux, Hélène and Démery, Vincent and Maggs, Anthony C.},
 journal = {The Journal of Chemical Physics},
 volume = {161},
 number = {18},
 year = {2024},
 month = {nov},
 pages={184504},
 issn = {1089-7690},
 publisher = {AIP Publishing}
}

@article{Harris1970,
 doi = {10.1007/bf01020444},
 title = {Hall coefficient in solution for a simple dynamical model},
 author = {Harris, S.},
 journal = {Journal of Statistical Physics},
 volume = {2},
 number = {4},
 pages = {379–385},
 year = {1970},
 issn = {1572-9613},
 publisher = {Springer Science and Business Media LLC}
}

@article{Friedman1968,
 doi = {10.1063/1.1664456},
 title = {Calculation of the Effect of Non-Brownian Motion on Some dc Transport Coefficients in Solution},
 author = {Friedman, Harold L. and Ben-Naim, Arieh},
 journal = {The Journal of Chemical Physics},
 volume = {48},
 number = {1},
 pages = {120–127},
 year = {1968},
 month = {jan},
 issn = {1089-7690},
 publisher = {AIP Publishing}
}

@article{Harris1972,
 doi = {10.1016/0009-2614(72)90015-2},
 title = {Hall coefficient in brownian-like solutions},
 author = {Harris, S.},
 journal = {Chemical Physics Letters},
 volume = {12},
 number = {3},
 pages = {493–494},
 year = {1972},
 month = {jan},
 issn = {0009-2614},
 publisher = {Elsevier BV}
}

@article{LaforgueKantzer1964,
 doi = {10.1051/jphys:01964002508-9084000},
 title = {Mise en évidence, sur les conducteurs ioniques, d’un effet magnétoélectrique semblable à l’effet Hall},
 author = {Laforgue-Kantzer, Denise},
 journal = {Journal de Physique},
 volume = {25},
 number = {8-9},
 pages = {840–842},
 year = {1964},
 issn = {0302-0738},
 publisher = {EDP Sciences}
}

@article{Chechetkin1989,
 doi = {10.1080/00268978900102681},
 title = {On the Hall number in dilute electrolyte solutions},
 author = {Chechetkin, V.R. and Lutovinov, V.S.},
 journal = {Molecular Physics},
 volume = {68},
 number = {4},
 pages = {979–981},
 year = {1989},
 month = {nov},
 issn = {1362-3028},
 publisher = {Informa UK Limited}
}

@article{Kroh1988,
 doi = {10.1016/0378-4371(88)90103-3},
 title = {On the theory of the hall effect in ionic solutions},
 author = {Kroh, H.J. and Felderhof, B.U.},
 journal = {Physica A: Statistical Mechanics and its Applications},
 volume = {153},
 number = {1},
 pages = {73–83},
 year = {1988},
 month = {nov},
 issn = {0378-4371},
 publisher = {Elsevier BV}
}

@article{Khanh1972,
 doi = {10.1016/0013-4686(72)85015-1},
 title = {Application de la statistique classique à l’étude de la conductibilité électrique et électromagnétique des solutions ioniques diluées},
 author = {Khanh, Tran Cong and Laforgue, A. and Laforgue-Kantzer, D.},
 journal = {Electrochimica Acta},
 volume = {17},
 number = {1},
 pages = {143–149},
 year = {1972},
 month = {jan},
 issn = {0013-4686},
 publisher = {Elsevier BV}
}

@article{zwanzig1973nonlinear,
  title={Nonlinear generalized Langevin equations},
  author={Zwanzig, Robert},
  journal={Journal of Statistical Physics},
  volume={9},
  number={3},
  pages={215--220},
  year={1973},
  publisher={Springer}
}

@book{elber2019molecular,
  title     = {Molecular Kinetics in Condensed Phases: Theory, Simulation, and Analysis},
  author    = {Elber, Ron and Orland, Henri and Makarov, Dmitrii E.},
  year      = {2019},
  publisher = {John Wiley \& Sons},
  }

@article{adar2018dielectric,
  doi = {10.1063/1.5042235},
  title={Dielectric constant of ionic solutions: Combined effects of correlations and excluded volume},
  author={Adar, Ram M and Markovich, Tomer and Levy, Amir and Orland, Henri and Andelman, David},
  journal={The Journal of chemical physics},
  volume={149},
  number={5},
  pages={054504},
  year={2018},
  publisher={AIP Publishing}
}

@article{Arato1962,
  title={Evaluation of the parameters of a complex stationary Gauss--Markov process},
  author={Arat{\'o}, M{\'a}ti{\'a}s and Kolmogorov, Andrei Nikolaevich and Sinai, Yakov Grigor'evich},
  journal={Dokl. Akad. Nauk SSSR},
  volume={146},
  number={4},
  pages={747--750},
  year={1962},
  url={https://www.mathnet.ru/eng/dan27047}
}

@article{Chen2014,
 doi = {10.1215/21562261-2693451},
 title = {On the eigenfunctions of the complex Ornstein–Uhlenbeck operators},
 author = {Chen, Yong and Liu, Yong},
 journal = {Kyoto Journal of Mathematics},
 volume = {54},
 number = {3},
 year = {2014},
 month = {jan},
 pages = {577--596},
 issn = {2156-2261},
 publisher = {Duke University Press}
}

@book{Balescu1997,
 doi = {10.1142/p036},
 title = {Statistical Dynamics: Matter Out of Equilibrium},
 author = {Balescu, Radu},
 year = {1997},
 month = {apr},
 isbn = {9781848160958},
 publisher = {Imperial College Press}
}

@article{Diosi1998,
 doi = {10.1103/physreva.58.1699},
 title = {Non-Markovian quantum state diffusion},
 author = {Diósi, L. and Gisin, N. and Strunz, W. T.},
 journal = {Physical Review A},
 volume = {58},
 number = {3},
 pages = {1699–1712},
 year = {1998},
 month = {sept},
 issn = {1094-1622},
 publisher = {American Physical Society (APS)}
}

@article{Lowen2016,
 doi = {10.1140/epjst/e2016-60054-6},
 title = {Chirality in microswimmer motion: From circle swimmers to active turbulence},
 author = {Löwen, Hartmut},
 journal = {The European Physical Journal Special Topics},
 volume = {225},
 number = {11-12},
 pages = {2319–2331},
 year = {2016},
 month = {nov},
 issn = {1951-6401},
 publisher = {Springer Science and Business Media LLC}
}

@article{Metafune2002,
 doi = {10.1006/jfan.2002.3978},
 title = {Spectrum of Ornstein-Uhlenbeck Operators in Lp Spaces with Respect to Invariant Measures},
 author = {Metafune, G. and Pallara, D. and Priola, E.},
 journal = {Journal of Functional Analysis},
 volume = {196},
 number = {1},
 pages = {40–60},
 year = {2002},
 month = {dec},
 issn = {0022-1236},
 publisher = {Elsevier BV}
}

@article{Kummel2013,
 doi = {10.1103/physrevlett.110.198302},
 title = {Circular Motion of Asymmetric Self-Propelling Particles},
 author = {Kümmel, Felix and ten Hagen, Borge and Wittkowski, Raphael and Buttinoni, Ivo and Eichhorn, Ralf and Volpe, Giovanni and Löwen, Hartmut and Bechinger, Clemens},
 journal = {Physical Review Letters},
 volume = {110},
 number = {19},
 pages = {198302},
 year = {2013},
 month = {may},
 issn = {1079-7114},
 publisher = {American Physical Society (APS)}
}

@article{Sevilla2016,
 doi = {10.1103/physreve.94.062120},
 title = {Diffusion of active chiral particles},
 author = {Sevilla, Francisco J.},
 journal = {Physical Review E},
 volume = {94},
 number = {6},
 pages = {062120},
 year = {2016},
 month = {dec},
 issn = {2470-0053},
 publisher = {American Physical Society (APS)}
}

@article{Kuroda2025,
 doi = {10.1103/25s2-6y3m},
 title = {Singular density correlations in chiral active fluids in three dimensions},
 author = {Kuroda, Yuta and Kawasaki, Takeshi and Miyazaki, Kunimasa},
 journal = {Physical Review E},
 volume = {112},
 number = {4},
 pages = {1103/25s2-6y3m},
 year = {2025},
 month = {oct},
 issn = {2470-0053},
 publisher = {American Physical Society (APS)}
}

@article{Jardat1999,
 doi = {10.1063/1.478703},
 title = {Transport coefficients of electrolyte solutions from Smart Brownian dynamics simulations},
 author = {Jardat, M. and Bernard, O. and Turq, P. and Kneller, G. R.},
 journal = {The Journal of Chemical Physics},
 volume = {110},
 number = {16},
 pages = {7993–7999},
 year = {1999},
 month = {apr},
 issn = {1089-7690},
 publisher = {AIP Publishing}
}

@article{Felderhof1983,
 doi = {10.1016/0378-4371(83)90111-5},
 title = {Linear response theory of sedimentation and diffusion in a suspension of spherical particles},
 author = {Felderhof, B.U. and Jones, R.B.},
 journal = {Physica A: Statistical Mechanics and its Applications},
 volume = {119},
 number = {3},
 pages = {591–608},
 year = {1983},
 month = {may},
 issn = {0378-4371},
 publisher = {Elsevier BV}
}

@article{Felderhof1987,
 doi = {10.1016/0378-4371(87)90278-0},
 title = {Linear response theory of the viscosity of suspensions of spherical brownian particles},
 author = {Felderhof, B.U. and Jones, R.B.},
 journal = {Physica A: Statistical Mechanics and its Applications},
 volume = {146},
 number = {3},
 pages = {417–432},
 year = {1987},
 month = {dec},
 issn = {0378-4371},
 publisher = {Elsevier BV}
}

\end{document}